\documentclass[
    ,final            
  ]
  {aipproc}

\layoutstyle{6x9}

\renewcommand\XFMtitleblock{
  \XFMtitle
  \let\XFMoldpar\par
  \def\par{\XFMoldpar\def\par{\space
	for the CTA Consortium\XFMoldpar}}
   \XFMauthors
   \let\par\XFMoldpar
   \XFMaddresses
   \XFMabstract
   \vspace{5pt}
   \XFMkeywords
   \XFMclassification
}

\begin{document}

\title{Towards a Flexible Array Control and Operation Framework for CTA}


\author{E. Birsin}{
  address={Institut für Physik, Humboldt-Universität zu Berlin }
}

\author{J. Colomé}{
  address={Institutde Cienciès de l'Espai (IEEC-CSIC)}
}

\author{D. Hoffmann}{
  address={Centre de Physique des Particules de Marseille}
  }

\author{H. Koeppel}{
  address={Deutsches Elektronen-Synchrotron, DESY}
}

\author{G. Lamanna}{
  address={Laboratoire d'Annecy-le-Vieux de Physique des Particules, Universit\'{e} de Savoie, CNRS/IN2P3, F-74941 Annecy-le-Vieux, France }
}

\author{T. Le Flour}{
	address={Laboratoire d'Annecy-le-Vieux de Physique des Particules, Universit\'{e} de Savoie, CNRS/IN2P3, F-74941 Annecy-le-Vieux, France }
}
\author{A. Lopatin}{
	address={Erlangen Centre for Astroparticle Physics (ECAP)}
}

\author{E. Lyard}{
	address={ISDC, Geneva Observatory, University of Geneva}
}

\author{D. Melkumyan}{
	address={Deutsches Elektronen-Synchrotron, DESY}
}

\author{I. Oya}{
	address={Institut für Physik, Humboldt-Universität zu Berlin}
}

\author{J-L. Panazol}{
	address={Laboratoire d'Annecy-le-Vieux de Physique des Particules, Universit\'{e} de Savoie, CNRS/IN2P3, F-74941 Annecy-le-Vieux, France }
}

\author{S. Schlenstedt}{
	address={Deutsches Elektronen-Synchrotron, DESY}
}

\author{T. Schmidt}{
	address={Deutsches Elektronen-Synchrotron, DESY}
}

\author{U. Schwanke}{
	address={Institut für Physik, Humboldt-Universität zu Berlin}
}

\author{C. Stegman}{
	address={Deutsches Elektronen-Synchrotron, DESY}
}

\author{R. Walter}{
	address={ISDC, Geneva Observatory, University of Geneva}
}

\author{P. Wegner }{
	address={Deutsches Elektronen-Synchrotron, DESY}
}

\begin{abstract}
The Cherenkov Telescope Array (CTA) \cite{CTA:2010} will be the successor to current Imaging Atmospheric Cherenkov Telescopes (IACT) like H.E.S.S., MAGIC and VERITAS. CTA will improve in sensitivity by about an order of magnitude compared to the current generation of IACTs. The energy range will extend from well below 100 GeV to above 100 TeV. To accomplish these goals, CTA will consist of two arrays, one in each hemisphere, consisting of 50-80 telescopes and composed of three different telescope types with different mirror sizes. It will be the first open observatory for very high energy $\gamma$-ray astronomy.\newline
The Array Control working group of CTA is currently evaluating existing technologies which are best suited for a project like CTA. The considered solutions comprise the ALMA Common Software (ACS), the OPC Unified Architecture (OPC UA) and the Data Distribution Service (DDS) for bulk data transfer. The first applications, like an automatic observation scheduler and the control software for some prototype instrumentation have been developed.
\end{abstract}

\maketitle


\section{Array Control and Operation}
The CTA instrument will have to deal with a large variety of observation modes and will operate with high efficiency in an automatic mode. This requires a well designed, failure tolerant system integrating both an automatic control layer and a user software layer for control and operation of the system. The Array Control and Operation (ACTL) group is responsible for instrument monitoring software, instrument slow control software, instrument operation software, data acquisition software, array trigger and online IT infrastructure development and maintenance.

\section{ALMA Common Software}
\begin{figure}
  \label{Fig:Control}
  \includegraphics[height=.3\textheight]{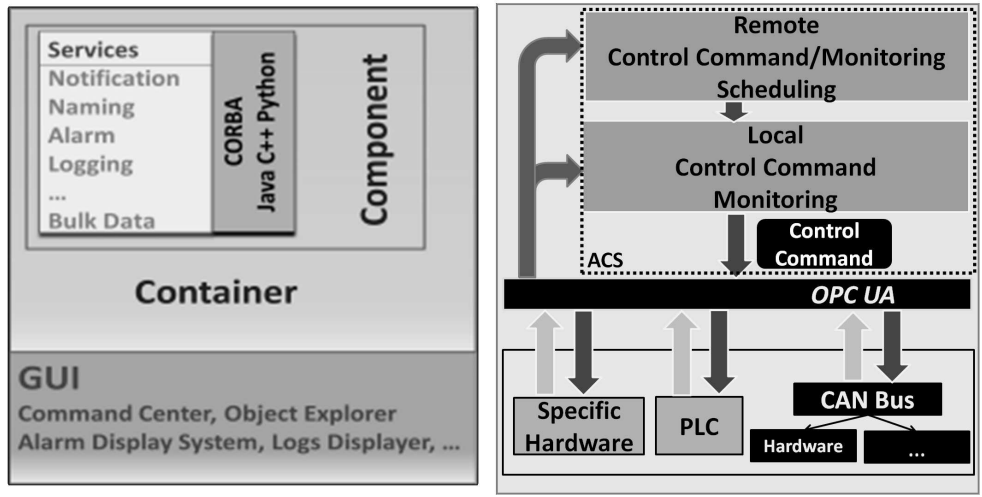}
  \caption{Left: ACS uses a container/component model. It provides important systems for array control like alarm system or logging service. Right: A thin OPC UA layer between ACS and hardware allows to contact different types of hardware with different connections without concrete knowledge of the connection.}
\end{figure}

As control software for CTA the ALMA Common Software (ACS) \cite{ALMA:2004} is considered. ACS was originally developed for the Atacama Large Millimeter Array (ALMA), an array of sixty millimeter to sub-millimeter antennas, and is still further developed by the European Southern Observatory (ESO).\newline
ACS is a CORBA based framework which follows a container/component model, see Fig. \ref{Fig:Control} left hand side. ACS supports C++, Java and Python. Important systems like alarm system and logging service is provided by ACS.\newline
Furthermore ESO provides a software component, Comodo, which allows the code generation from Unified Modelling Language (UML) diagrams, speeding up the process of code development.

\subsection{OPC Unified Architecture}
One key point in the software design is easy maintainability. Hardware parts should be replaceable with a minimum effort put into reprogramming software. Therefore hardware must be accessible in an unified way.\newline 
OPC Unified Architecture (OPC UA) \cite{opcuaweb} provides a set of defined standards making it possible to implement communication to different devices with different connections in a standardized way. The current design foresees OPC UA as a thin middle layer between hardware and ACS, see Fig.\ref{Fig:Control} right hand side. \newline
For OPC UA server/client programming in C++ and Java software development kits from Unified Automation Company and Prosys, respectively, have been chosen. First server and client components are running. The design will be tested on the medium sized prototype telescope (MST) \cite{Oya:2012}.

\subsection{Data Distribution Service}
A total data rate ranging from 300 MB/s up to 4 GB/s is expected \cite{Arribas:2012} for both CTA sites. To handle such high data rates and also assure data quality, the use of Data Distribution Service (DDS) is considered. DDS works with a publish/subscribe model for distributed systems. Performance tests of several implementations of DDS such as RTI DDS, OpenSplice DDS and openDDS are carried out.

\section{Observation Scheduler}
The observation scheduler has to optimise the usage of available observation time. Many factors are considered like visibility, science priority, weather conditions and slewing time. The scheduler is also taking into account that CTA can be split into sub-arrays which will operate in parallel on different observation targets.\newline
To optimise the amount of data relevant for physics the scheduler is foreseen to have two parts. A short-term and a long-term scheduler. The short-term scheduler will be able to react to targets of opportunity, like gamma-ray bursts, and dynamically adjust the observation schedule to current conditions. The long-term scheduler on the other hand is responsible for the scheduling from longer term like night or even season planning.


\begin{theacknowledgments}
We gratefully acknowledge support from the agencies and organizations 
listed in this page: \url{http://www.cta-observatory.org/?q=node/22}

\end{theacknowledgments}



\bibliographystyle{aipproc}   


\IfFileExists{\jobname.bbl}{}
 {\typeout{}
  \typeout{******************************************}
  \typeout{** Please run "bibtex \jobname" to optain}
  \typeout{** the bibliography and then re-run LaTeX}
  \typeout{** twice to fix the references!}
  \typeout{******************************************}
  \typeout{}
 }

\end{document}